\documentclass[11pt]{article}
\usepackage{epsfig}
\begin{document}
\pagestyle{empty}
\begin{center}
{\Large Contribution of real and virtual Pions to nuclear Photoabsorption
at Intermediate Energies\footnote{Work supported by Deutsche
Forschungsgemeinschaft
(contracts Schu 222/17 and 438/113/173)}
 } \\
\vspace{0.5cm}
M.-Th.\,H\"utt$^a$, A.I.\,Milstein$^b$ and M.\,Schumacher$^a$ \\
 \footnotesize
       (a) II. Physikalisches Institut der Universit\"at G\"ottingen,
       G\"ottingen, Germany

       (b) Budker Institute of
       Nuclear Physics, 630090 Novosibirsk, Russia
\end{center}
\date{}

\parindent0em
\parskip1.5ex plus 0.5ex minus 0.5ex

\sloppy
\renewcommand{\textfraction}{0.05}
\renewcommand{\topfraction}{0.99}
\renewcommand{\bottomfraction}{0.99}
\newtheorem{abb}{Figure}
\newcommand{\bi}[1]{\bibitem{#1}}

\begin{abstract}
The universal curve $\sigma /A$ of nuclear photoabsorption is
investigated within a Fermi gas model of nuclear matter. An energy range
from pion threshold up to 400 MeV is considered. The interactions between
nucleon, pion, $\Delta $-isobar and photon are considered
in the non-relativistic approximation with corrections of the order
$1/M$ taken into account with respect to proton mass.
Analytical expressions are obtained, in which the influence of
nuclear correlations, two-nucleon contributions and relativistic
corrections is studied explicitely. An extension of the model
calculation to nucleon knock-out  reactions is discussed.
\end{abstract}



\newpage
\pagestyle{plain}
\pagenumbering{arabic}
\section{Introduction}
The phenomenology of nuclear photoabsorption is governed by two
characteristic features. First, all nuclei with mass numbers $A$ ranging
from 10 to more than 200 obey the same fundamental curve
$\sigma  (\omega )/A$ for the total photoabsorption cross section
devided by $A$ as a function of the photon energy $\omega $. Second, the
$\Delta $-isobar excitation of the nucleon is responsible
for the main properties of this curve in the energy region between 200
and 400 MeV. Models, which focus on the behaviour of the $\Delta $-isobar
in a nuclear environment, namely the $\Delta $-hole calculations 
\cite{bb,cc}, have
proven to be highly successful in explaining the experimental findings for
pion scattering processes \cite{aa}.
Indeed, on grounds of the $\Delta $-hole formalism a wide variety of 
pion-nucleus reactions can be described within one consistent framework 
\cite{dd}.
In the case of photonuclear reactions, however, serious descrepancies
remain, which partially have been accounted for by including
non-resonant background terms \cite{koch,ee}. Nevertheless, such a
procedure, in particular for nuclear photoabsorption, either leads to
contradictions with previous $\Delta $-hole results or lacks the complete
agreement with experimental data \cite{koch}. The question arises, whether
theoretical ingredients other than in-medium $\Delta $-hole propagations
can lead to a similar degree of accuracy. Therefore it is natural to
address the situation from a different point of view asking to what
extend the absorption process can be described, when only a very simple
$\Delta $-nucleon interaction is used
and all additional effects are accounted for in a purely diagrammatic
approach.
When combined with a simple form of nucleon momentum distribution inside
the nucleus, namely a Fermi gas model, this leads to analytical
expressions, in which different corrections to this lowest-order
approximation can be studied. This is the aim of the present article.
In our formalism we follow closely Wakamatsu and Matsumoto \cite{waka}.
However, we do
not introduce a phenomenological potential to account for the
deviation of the nucleon wave functions from plane waves, but study the
influence of such corrections in a perturbative way, similar to
our previous work \cite{hm}.
In addition, our focus is on the energy-dependence of the total
photoabsorption,
rather than on the differential cross section as a function of the
momentum of the outgoing proton. 
A characteristic feature of Wakamatsu's and Matsumoto's 
approach is the technically equal 
treatment of the ($\gamma $,pn) and the ($\gamma $,pp) knock-out process, 
which allowed them to obtain a natural explanation for the supression of 
the two-proton knock-out. This characteristic property is also present 
in our calculation, where it is related to a vanishing trace in spin space.
In the last years the ratio ($\gamma $,pp)/($\gamma $,pn) has been 
investigated in depth within different formalisms. In an extension of 
Wakamatsu's and Matsumoto's work by Boato and Giannini \cite{ff}, 
finite-size effects have been calculated and, more recently, a 
combination of pion exchange and shell-model wave functions was used to 
investigate this quantity \cite{ryck1}.
Currently, two complementary approaches for the description of nuclear 
knock-out reactions exist. Carrasco and Oset \cite{oset1} used a 
diagram-oriented many-body expansion in a Fermi gas. The evaluation of
self-energy diagrams leads to an accuracy for medium effects high enough 
to study knock-out reactions in great detail. 
With their primary goal being thus different from ours, their formalism
does not yield isolated expressions for the resonant and non-resonant 
parts of the mechanisms of nuclear photoabsorption
A different approach is used by the Gent group 
\cite{ryck2,ryck4,ryck5}, where the
main emphasis lies in the construction of realistic shell-model wave functions, 
rather than on a microscopic description fully based on the evaluation of 
Feynman diagrams. As the nuclear photoabsorption is almost insensitive to 
structural differences between nuclei, the quality of their approach 
becomes obvious in the investigation of differential cross sections for 
nucleon knock-out, rather than of photoabsorption.

Nuclear photoabsorption provides an interesting tool to study the
interplay between one-nucleon and two-nucleon contributions.
We obtained analytical expressions for these contributions, as well as 
for their resonant and non-resonant parts. This set of results 
can be used as a starting point to include (and test) further nuclear or nucleonic
effects.
In Section 2 the basic notations are listed, as well as the
interaction terms and the most important model assumptions and
approximations, which are present in this
calculation. The main results and their most prominent properties, e.g.
the effects of relativistic corrections and nuclear structure,
are discussed in Section 3, where also the following
possible extension of such an approach is considered: 
As the angular dependence of the
two-nucleon process is not very strong (cf. \cite{angle}), the different 
mechanisms contributing to the photoabsorption curve can also be used to 
understand qualitative features of experimental data for nucleon knock-out 
processes as a function of the photon energy.
In Section 4 some
concluding remarks are made, with an emphasis on the applicability of the
partial cross sections, whose analytical forms are given in the Appendix.

\section{Formalism and Notation}
The starting point of our investigation is the static Hamiltonian for the
pion-nucleon interaction,
\begin{equation}
        H_{\pi NN}=-{f \over m}\;\vec \sigma \cdot \vec \nabla \;\underline
        \tau \cdot \underline \phi
        \label{s1eq1}
\end{equation}
together with a minimal coupling to the photon field. In eq.\,(\ref{s1eq1})
$m$ is the pion mass.
For all coupling constants we use the same notation and values as given in
\cite{ew}, in particular $f^2/(4\pi )$=0.08.
In eq.\,(\ref{s1eq1}) underlined symbols denote vectors in (cartesian) isospin
space, while an arrow indicates a vector in coordinate space.
In the static limit the interactions with the $\Delta $-isobar excitation
of the nucleon are determined by the following Hamiltonians (see e.g.
\cite{ew}):
\begin{equation}
        H_{\gamma N\Delta }=-{{ef_{\gamma N\Delta }}
        \over m}\,\vec S^+\cdot \left( {\vec \nabla \times \vec A} \right)\,\,T_z^+
        \label{s1eq2}
\end{equation}
and
\begin{equation}
        H_{\pi N\Delta }=-{{f_\Delta } \over m}\,\vec S^+\cdot
        \vec \nabla \underline T^+\underline \phi  \,\, ,
        \label{s1eq3}
\end{equation}
with the hermitian conjugate to be added in both cases. Here
$\vec S$ and $\underline T$ are the 1/2-to-3/2 transition operators in
spin space and isospin space, respectively;
$e$ is the proton charge, $e^2=1/137$.
For the coupling  constants we have $f_\Delta $=2 and $f_{\gamma N\Delta
}$=0.35. In all cases, where high momentum transfers
occur at the pion-nucleon vertices we regularize the vertex functions by
introducing dipole form factors
\begin{equation}
g_{\pi}(q)={{\Lambda ^2-m^2} \over {\Lambda ^2- q^2}}
        \label{s1eq4}
\end{equation}
as was also done e.g. in \cite{waka,Riska}. The value for the cut-off parameter
$\Lambda $ has been taken to be 800 MeV. This value  gives the best
agreement of our predictions with the experimental data. In addition,
a similar value has been used in \cite{sasa,waka}.
The general expression for
the total cross section $\sigma_1(\omega)$ of the photoabsorption
with one nucleon outside the Fermi sphere and one pion in the final state
(one-nucleon  process) is of the form
\begin{eqnarray}
& &\sigma _1(\omega )=\int {{{V\,d\vec p}
\over {(2\pi )^3}}}\,\int {{{d\vec q} \over {2\varepsilon _q(2\pi
)^3}}}\,{{4\pi }\over {2\omega }}\;\left| {T_1} \right|^2\times
\nonumber \\
& &2\pi \,\delta \left( {\omega +{{p^2}
\over {2M}}-\varepsilon _q-{{(\vec k+\vec p-\vec q)^2} \over {2M}}}
\right)\;n(\vec p)\;\left[ {1-n(\vec k+\vec p-\vec q)}\right]\; .
\label{s1eq5}
\end{eqnarray}
For the total cross section $\sigma_2(\omega)$ of the process
with two free nucleons in the final state it is given by
\begin{eqnarray}
& &\sigma _2(\omega )=\int {{{V\,d\vec p_1\,V\,d\vec p_2}
\over {(2\pi )^6}}}\,\int {{{d\vec p_3\,d\vec p_4\,} \over {(2\pi
)^6}}}\,{{4\pi }
\over {2\omega }}\;\left| {T_2} \right|^2\,\times
\nonumber \\
& &\delta \left( {\omega +{{p_1^2+p_2^2-p_3^2-p_4^2}
\over {2M}}} \right)\,(2\pi )^4\,\delta (\vec p_1+\vec p_2+\vec
k-\vec p_3-\vec p_4)
\times
\nonumber  \\
    & &n(\vec p_1)n(\vec p_2)\left[ {1-n(\vec p_3)\;} \right]\left[
{1-n(\vec p_4)\;}
    \right]
        \label{s1eq6}\; .
\end{eqnarray}
The notation for the external momenta is shown in Fig.\,(\ref{figA}).
The function $n(\vec p)=\theta (p_F-\left| {\vec p}\right|)$ is
the occupation number, $\theta(x)$ is the step function. Furthermore, $V$
is the nuclear volume, $p_F$ is the Fermi momentum, $M$ is the mass of
the proton, $V$ is the nuclear volume, $V=3\pi ^2A/(2p_F^3)$,
and $\varepsilon _q=\sqrt {\left| {\vec q} \right|^2+m^2}$ is the energy of
the outgoing pion.

Both amplitudes $T_1$ and $T_2$ consist of non-resonant and
resonant parts,
$T_i=T_i^{(NR)}+T_i^{(R)}$. Diagrammatically this decomposition is
shown in Fig.\,(\ref{figB}). The second (crossed) contribution to the
resonant part is small due to the big energy denominator and will be
skipped in the following. In the energy $\delta $-function in
eq.\,(\ref{s1eq5}) and eq.\,(\ref{s1eq6}) we will usually not consider the
smallest term connected with the kinetic energy of the incoming nucleon,
as it is smaller than $p_F^2/2M\approx 37\,$MeV. Indeed, as we expect its
contribution to introduce only a small modification of the actual
$p$-dependence in the integrand in (\ref{s1eq5}), we substitute it by
its average value
$\left\langle {p^2} \right\rangle /2M=(3/5)p_F^2/2M$, which results in
an overall shift of the absorption cross section. For the one-nucleon
process we find first-order relativistic corrections to be essential for
a successful treatment of the absorption process. In the case of the
resonant contribution, such corrections are accounted for by making the
following substitutions in the vertices \cite{waka}:
\begin{eqnarray}
&\vec q&\buildrel {} \over \longrightarrow \;\vec q-{{\varepsilon _{\vec q}}
\over {M_\Delta }}(\vec p+\vec k)  ,
\label{s1eq7} \\
&\vec k&\buildrel {} \over \longrightarrow \;\vec k\left( {1+{\Delta
\over M}}\right)-{\Delta  \over M}\,\vec p \, ,
\label{s1eq8}
\end{eqnarray}
where $M_\Delta $ is the $\Delta$-isobar mass and $\Delta=M_\Delta -M$.
For the non-resonant part, the corrections
give amplitude $T_1$ of the following form:
\begin{eqnarray}
        T_1^{(NR)}&=&{{\sqrt 2ef} \over m}\,\left\{ {\matrix{{}\cr
{}\cr
}} \right.i\,\vec \sigma \cdot \vec \varepsilon \,\left[ {1+{{\varepsilon _q}
\over {2M}}} \right]-{{2i(\vec \sigma \cdot (\vec q-\vec k))(\vec
\varepsilon \cdot \vec q)}\over {(\vec q-\vec k)^2+m^2-(\varepsilon
_q-\omega )^2}} \nonumber \\
&-&{{2i(\vec \sigma \cdot \vec q)(\vec \varepsilon \cdot (\vec k+\vec
p-\vec q))}\over {\,\left[ {(\vec k+\vec p-\vec q)^2-2M\omega -(\vec
p-\vec q)^2}\right]}}\left. {\matrix{{}\cr
{}\cr
}} \right\}  ,
        \label{s1eq9}
\end{eqnarray}
which corresponds to the production of a $\pi ^-$.
In comparison with \cite{waka} we neglected those terms in
(\ref{s1eq9}), which modify  the result by less than 2 per cent.
It should be noted that no free parameters are present in our approach. A
certain model dependence exists, however, in the selection of diagrams.
We neglect all those diagrams, which are suppressed by some
mechanism. In Fig.\,(\ref{figAA}) two examples for suppression
mechanisms are  given. For Fig.\,(\ref{figAA}a) the contribution is
small because in the case of infinite nuclear matter due to momentum
conservation the four-momentum of the photon should be equal to that of
the outgoing pion, which is impossible. For Fig.\,(\ref{figAA}b) let us
consider the case, where the upper two nucleons (incoming and
outgoing) are identified. Furthermore, let us select a $\Delta $-isobar
as an intermediate state. Then, one has a vanishing 
trace in spin space:
$$
\mbox{Sp}\left[ (\vec S^+\cdot (\vec k\times \vec \varepsilon
))\;(\vec S\cdot \vec q)\right]\,=\, 0
$$
at $\vec q\,=\vec k$.
Therefore, the contribution in this case
vanishes in the non-relativistic limit considered here.
As a result of applying such methods to the various diagrams, we have
obtained that  only those displayed in
Fig.\,(\ref{figA}) should be taken into acount.

\section{Results}
Explicit evaluation of the diagrams shown in Fig.\,(\ref{figA}) leads
to amplitudes  for the one- and two-nucleon contribution to nuclear
photoabsorption.  Squaring these amplitudes and summing over spin and
isospin states of the  nucleons by using trace methods as previously
\cite{hm} one finds the expressions
$\left| {T_1} \right|^2$ and $\left| {T_2} \right|^2$, which enter into
eqs.\,(\ref{s1eq5}) and (\ref{s1eq6}).
It has turned out to be convenient to investigate the resonant and
non-resonant parts of each of these contributions separately. This can be
done by neglecting the interference terms, which are highly suppressed
(cf. Fig.\,(\ref{figH})).
Shown here exemplary for the resonant parts, one obtains the following
expressions, which serve as a starting point for the integrations
with respect to  nucleon momenta:
\begin{eqnarray}
        & &\sigma _1^{(R)}(\omega )={{4e^2} \over {27\pi
        \omega }}\,\left( {{{f_\Delta f_{\gamma N\Delta }}
        \over {m^2}}} \right)^2\int {{{V\,d\vec p}
        \over {(2\pi )^3}}}\,\int {d\vec q}\;\times
        \nonumber \\
        & &\delta \left( {\omega ^2+{\omega  \over M}\left[
         {(\vec p-\vec k)^2-(\vec p-\vec q)^2}
        \right]-q^2-m^2} \right)\times
        \nonumber \\
        & &n(\vec p-\vec k)\;\left[ {1-n(\vec p-\vec q)}
\right]\,{{3\left[ {(\vec Q\times
        \vec K)\cdot \vec \varepsilon } \right]^2+Q^2(\vec K\times
\vec \varepsilon )^2}
        \over {\left( {\omega -\Delta - p^2/2M} \right)^2+
        \Gamma^2 / 4}}  ,
        \label{s2eq1}
\end{eqnarray}
where
\begin{displaymath}
        \vec Q=\vec q-{\omega  \over {M_\Delta }}\,\vec p\;,\quad
\vec K=\vec k\left( {1+{\Delta\over M}} \right)-{\Delta  \over M}\,\vec p
\end{displaymath}
and
\begin{eqnarray}
        & &\sigma _2^{(R)}(\omega )={{16e^2} \over {81\pi }}\,\left(
{{{f_\Delta f_{\gamma N\Delta }}\over {m^3}}} \right)^2\,{\omega
\over {\left( {\omega -\Delta } \right)^2+{{\Gamma ^2}/4}}}\times
        \nonumber \\
        & &\int {{{V\,d\vec p_1\,V\,d\vec p_2}
        \over {(2\pi )^6}}}\,\int {d\vec p_3\,d\vec p_4\,}n(\vec
p_1)n(\vec p_2)\left[
        {1-n(\vec p_3)\;} \right]\left[ {1-n(\vec p_4)\;} \right]\times
        \nonumber \\
        & &\delta \left( {\omega -{{p_3^2+p_4^2}
        \over {2M}}} \right)\,\delta (\vec p_1+\vec p_2+\vec k-\vec
p_3-\vec p_4)\times
        \nonumber \\
        & &\left\{ {{{2a^2\left[ {a^2+3(\vec \varepsilon \cdot \vec
a)^2} \right]}
        \over {(a^2+m^2)^2}}\,g_\pi ^4(a)+{{2\omega ^2(\vec \varepsilon \cdot
        \vec a)(\vec \varepsilon \cdot \vec b)}
        \over {(a^2+m^2)(b^2+m^2)}}\,g_\pi ^2(a)g_\pi ^2(b)} \right\}
        \label{s2eq2}
\end{eqnarray}
with
\begin{displaymath}
        \vec a=\vec p_4-\vec p_2\;,\quad \vec b=\vec p_1-\vec p_3  .
\end{displaymath}
The analytical expressions for the resulting partial
cross sections are given in the Appendix. For the total absorption cross
section $\sigma (\omega )$ a comparison with experimental data is
shown in Fig.\,(\ref{figC}). 
The result of this model calculation compares favorably with the data.
The one-nucleon and two-nucleon parts of the
cross section are equally important at energies around 250 MeV. As can be 
seen in this plot, significant features of the data, e.g. the position  of 
the peak, are only obtained due to the interplay between the two 
mechanisms. 
As mentioned before, each of the contributions to the full curve in 
Fig.\,(\ref{figC}) has a resonant and a non-resonant part. 
In Fig.\,(\ref{figD})
and (\ref{figE}) this decomposition is shown for the one-nucleon and the
two-nucleon mechanism, respectively. Here it is clearly seen that the 
non-resonant parts give an important contribution at lower energies. In 
the two-nucleon case the non-resonant part decreases with energy, while 
in the one-nucleon process it remains almost constant. 

It is interesting to see, in what way the one-nucleon partial cross
sections are affected by the use of only the static limit of the
interaction.
Neglecting the first-order relativistic
corrections in the current one has
\begin{eqnarray}
        \tilde \sigma ^{(R)}_1(\omega )={{Ae^2(f_\Delta f_{\gamma
N\Delta })^2}
        \over {27m^2\left[ {\left( {\omega -\Delta } \right)^2+{{\Gamma ^2}
        \over 4}} \right]}}\,\int\limits_{q_{min}}^{q_{max}}
{q\,dq}\,\left( {1+S_0(q)}  \right)\times  & &
        \nonumber \\
        \left\{ {\left[ {\omega ^2-m^2-{{\omega q^2}
        \over M}} \right]+{3 \over 2} \left( {q^2-\left[ {q^2\left(
{1+{\omega \over M}} \right)+m^2} \right]^2{1 \over {4\omega ^2}}} \right)}
        \right\}  & &
        \label{s2eq3}
\end{eqnarray}
and
\begin{eqnarray}
         & &\tilde \sigma ^{(NR)}_1(\omega )={{Ae^2f^2}
         \over {m^2\omega ^2}}\,\int\limits_{q_{min}}^{q_{max}}
{q\,dq}\,\left( {1+S_0(q)}
         \right)\left[ {\matrix{{}\cr
{}\cr
}} \right.1-{{2m^2q^2} \over {(q^2+m^2)^2}}\times
        \nonumber \\
         & &\left\{ {1-{1 \over {4q^2\omega ^2}}\left[ {q^2\left( {1+{\omega
         \over M}} \right)+m^2} \right]^2} \right\}\left. {\matrix{{}\cr
{}\cr
}} \right]  ,
        \label{s2eq4}
\end{eqnarray}
where the integration limits in both cases are given by
\begin{equation}
        q_{\min ,\max }=\left( {1+{\omega  \over M}}
        \right)\,\left\{ {\omega \mp \sqrt {\omega ^2-m^2\left( {1+{\omega
        \over M}} \right)\,}} \right\}
        \label{s2eq5}
\end{equation}
and the integrand contains the function
\begin{equation}
        S_0(q)=-\left( {1-{q \over {2p_F}}} \right)^2\left( {1+{q
        \over {4p_F}}} \right)\,\theta (2p_F-q)  .
        \label{s2eq6}
\end{equation}
In Fig.\,(\ref{figF}) the corresponding cross sections are compared with those
resulting from eq.\,(\ref{s2eq1}) and its non-resonant counterpart.
The relativistic corrections lead mainly to a shift of the one-nucleon 
curve. This effect is essential for obtaining a good agreement with the 
experimental data. 
Although no relativistic corrections in the current are included, note
that in (\ref{s2eq3}) and (\ref{s2eq4}) terms of that order have
been kept in the kinematical contributions to the integrand.
As the two-nucleon mechanism gives a comparatively small contribution at 
higher energies, we neglect relativistic corrections in $\sigma 
_2(\omega )$. This has also been done in \cite{waka}.

A more difficult problem is the influence, which nucleon correlations inside
the nucleus can have on the nuclear photoabsorption process.
We investigate this aspect for the
non-relativistic forms (\ref{s2eq3}) and (\ref{s2eq4})
of the one-nucleon case. The main reason for
doing so is the fact that due to the square of the amplitudes we can
express part of the integrand via the standard lowest-order central
correlation function of a Fermi gas (see e.g. \cite{Molinari}). The
object, which is
obtained by diagrammatically squaring the one-nucleon contribution of
Fig.\,(\ref{figA}a), can be coupled to an additional nucleon. In the
incoherent case of the diagrammatical square also a
further pion exchange can be allowed. The modifications of the
correlation function, which occur due to such effects, have been
investigated analytically in \cite{hm}, where a corrected 
central correlator $S_C(q)$ has been constructed. 
The function
$S_C$ is given in Fig.\,(\ref{figJ}).
By making the substitution
$ S_0(q)\buildrel {} \over \longrightarrow S_C(q) $
these further correlations can be incorporated effectively.  
In Fig.\,(\ref{figG}) the 
one-nucleon contribution resulting from $S_C$ is compared with the 
original form, in which $S_0$ has been used. It can be seen that such medium 
effects modify the result by about 15 per cent. Naturally, the use of this
substitution technique can only be used to obtain an estimate for such a 
medium-induced modification. A full examination should involve the 
inclusion of the additional two-and three-nucleon diagrams 
in eq.\,(\ref{s1eq5}).

The limit of our approach is certainly reached, when a comparison with
differential cross sections is attempted.
An extreme case is the comparison with $^4$He, for which a recent
measurement of both, the one-nucleon and the two-nucleon channel exists
\cite{wich}. 
We compare the calculated average cross section with the data for the 
differential cross section in c.m. frame, as we expect the angular 
dependence not to be strong in that frame. The
corresponding plot is shown in Fig.\,(\ref{figI}). 
Although no full agreement is obtained, it is interesting to note that 
the general features of the two cross sections are well reproduced, such 
as the peak positions and the relative size of the two processes. By 
these means it is possible to unambiguously identify the physical 
mechanisms behind the data points.
A similar degree of agreement is obtained for other data \cite{homma2}.

\section{Conclusion}
In the present paper we have developed a diagrammatical description of 
the nuclear photoabsorption process. The main result of our investigation 
is that the total photoabsorption cross section can be fully understood in 
terms of a simple physical picture, where point-like nucleons and $\Delta 
$-isobars interacting via pion exchange are the relevant degrees of 
freedom. Due to the diagram-oriented formalism and the Fermi gas model as 
an approximate description of the nucleons in momentum space, 
we could obtain analytical 
expressions for all the relevant contributions to the photoabsorption 
curve. In this way a flexible and efficient description has been 
obtained, which can be used as a starting point for the investigation of 
additional effects.
Especially in the low-energy part of our calculated curve, the
agreement with experiment comes about as a non-trivial interplay
between the one-nucleon and two-nucleon contributions.
It is worth noting that, as long as a comparatively low cut-off parameter 
in the vertex form factor is used, there seems to be no need for an explicit
diagrammatical inclusion of the $\rho $-meson as an additional mechanism
of the nucleon-nucleon interaction. 
We found relativistic corrections in the case of the one-nucleon
process to be crucial for obtaining a good agreement.
The aspect of additional nucleon correlations, which can be accounted for 
as a deviation of the nucleon wave functions from plane waves, deserves 
some further attention in future investigations. We could estimate the 
overall effect to be of the order of 15 per cent.

\section*{Acknowledgement}
We are most grateful to A.I. L'vov for useful comments and
discussions. One of us (M.T.H.) wishes to thank the Budker
Institute, Novosibirsk, for the
kind hospitality accorded him during his stay, when part of this work was
done.

\section*{Appendix: Analytical expressions for absorption cross sections}

Here we present the explicit expressions for the four contributions to
the photoabsorption curve,
which have been used to obtain the figures shown in Section 3.
As was mentioned earlier, the interference terms between the resonant and
the non-resonant contributions are small (cf. Fig.\,(\ref{figH})).
Therefore, we can
write the absorption cross section as a sum of four parts,
 $\sigma (\omega )=\sigma _1^{(NR)}(\omega )+
\sigma _1^{(R)}(\omega )+\sigma _2^{(NR)}(\omega )+\sigma
_2^{(R)}(\omega )$.
First, we deal with the one-nucleon case.
In the case of the non-resonant contribution the absorption cross
section can be represented in the following form:
\begin{eqnarray}
&\displaystyle
    \sigma _1^{(NR)}(\omega )=
\frac{3Ae^2}{4m^2\omega ^2p_F^3}\,
\int\limits_{\omega_{-}}^{\omega_{+}} dq\int\limits_{l(q)}^
    {q+p_F} dp\,g_\pi^2(q)\,p\,[p_F^2-(q-p)^2]\times \nonumber\\
&\displaystyle
G(p,q,\omega)\,\theta(2q\omega-m^2-q^2-\frac{\omega}{M}p^2) ,
        \label{app1}
\end{eqnarray}
where
\begin{equation}
G(p,q,\omega )\!=\!\left(1+\!\frac{\omega}{2M}\right)^2\!-\!
\frac{2}{D}\left(\frac{m^2}{D}+\frac{\omega}{2M}\right)
\left[q^2-{1 \over {4\omega ^2}}\left( D+{{\omega}\over M}p^2\right)^2
\right]
        \label{app4}
\end{equation}

and $l(q)=\mbox{Max}[p_F,|q-p_F|]$ , $D=q^2+m^2$ ,
$\omega_{\pm}=\omega \pm\sqrt{\omega ^2-m^2}$.
In eq.\,(\ref{app1}) the form factor $g_\pi ^2(q)$ is the same as in
(\ref{s1eq4}). Note that the integration with respect to the variable $p$
can easily be performed, but the result is too lengthy to be given here 
explicitely.
The resonant part of the one-nucleon process has the following form:
\begin{eqnarray}
&\displaystyle
        \sigma _1^{(R)}(\omega )=
\frac{4Ae^2}{9\omega p_F^3}\left(\frac{f_{\gamma N\Delta }}
{m^2}\right)^2\left(1+{\Delta\over M}\right)^2
\left(1+{\omega\over M}\right)^2 \int\limits_{-1}^{+1} {dy}
\int\limits_0^b dp\, p^2F\times \nonumber \\
&\displaystyle
\frac{\theta \left(p_F^2-p^2\left(1+\omega/M \right)^2
-\omega ^2+2\omega py\left(1+\omega/M \right)\right)}
{\left[\omega -\Delta -p^2(1+\omega/M)^2/2M_{\Delta}\right]^2+
\Gamma^2/4}\; H(p,y,\omega)
\label{app6}
\end{eqnarray}
where $b=(\omega+p_F)/(1+\omega/M)$ and the integrand is given by
\begin{eqnarray}
&\displaystyle
H(p,y,\omega)=
2A_1\, [\theta(F-p-p_F)+\theta(F)\,\theta(p-p_F-F)\,]+
\nonumber\\
&\displaystyle
[(x+1)A_1-{1 \over 8}(x-x^3)A_2]\theta(F-|p-p_F|)\,\theta(p+p_F-F)
     \, ,
\end{eqnarray}
with
\begin{eqnarray}
&\displaystyle
F=\left[\omega^2\left(1-\frac{2py}{M}\right)
-\frac{m^2}{1+\omega/M}\right]^{1/2}\; ,
\; x=\frac{p^2+F^2-p_F^2}{2pF}\, ,  \nonumber \\
&\displaystyle
A_1=\omega^2-2\omega ay+a^2\; ,\; A_2=\omega^2(1-3y^2)+4\omega ay-2a^2\; ,
\nonumber \\
&\displaystyle
a=\frac{p\Delta}{M}\left(\frac{1+\omega/M}{1+\Delta/M}\right)\, .
\end{eqnarray}
The mass difference $\Delta $ between the proton and the $\Delta
$-excitation is $\Delta=292$ MeV, the width $\Gamma $ of the 
$\Delta$-isobar has been taken to be 115 MeV.
Again, the integration with respect to $y$ in eq.\,(\ref{app6}) can be performed
analytically, but due to its length the result is not presented here.
For the partial cross sections of the two-nucleon case, we obtained the
following result:
\begin{eqnarray}
&\displaystyle
\sigma _2^{(NR)}(\omega )+\sigma _2^{(R)}(\omega )=\\
&\displaystyle
\Biggl\{ \theta (\omega -\varepsilon )\,\theta (5\varepsilon -\omega)
\,\Biggl[\,\int\limits_{L_1(Q)}^{L_2(Q)} \Phi(\beta_1,\, Q)+
\int\limits_{L_2(Q)}^2 \Phi(\beta_2,\, Q)\Biggr]+\nonumber \\
&\displaystyle
\theta(\omega -5\varepsilon )\,\theta (9\varepsilon -
\omega )\,\int\limits_{L_1(Q)}^{L_3(Q)}\Phi(\beta_1,\, Q)
\Biggr\}g_\pi ^4(p)\left[G^{(NR)}(p,\omega )+G^{(R)}(p,\omega )\right]dp ,
\nonumber
        \label{app10}
\end{eqnarray}
where $Q=\sqrt {M\omega }$ , $\varepsilon =p_F^2/M$ ,
$$
\beta_1=\arccos \frac{p+p_F}{\sqrt{2}Q} \; ,
\beta_2=\arccos \frac{p_F}{\sqrt {2}Q}\; .
$$
The elementary function $\Phi(\beta,\, Q)$ is
\begin{eqnarray}
&\displaystyle
\Phi(\beta,\,Q)=
\Biggl\{ \frac{1}{2}(p^2-p_F^2)(2Q^2+p^2-p_F^2)\sin ^2x\,+Q^4
(\sin ^4x-\frac{2}{3}\sin ^6x)\, +  \nonumber \\
&\displaystyle
p^2Q^2(x-\frac{1}{4}\sin 4x\,)
-\sqrt 8pQ\, [\frac{1}{3}(2Q^2+p^2-p_F^2)\sin ^3x-
\frac{2}{5}Q^2\sin ^5x\,]\, +  \nonumber \\
&\displaystyle
\sqrt 8pQ\,[\frac{1}{3}(2Q^2+p^2-p_F^2)\cos ^3x
     -\frac{2}{5}Q^2\cos ^5x\,]\Biggr\}\,
  \Biggl.\Biggr|_{x=\beta}^{x=\pi /2 -\beta}\, .
        \label{app13}
\end{eqnarray}
The functions in the integrand of eq.\,(\ref{app10}), which characterize
the resonant and non-resonant part, are given by
\begin{equation}
G^{R}(p,\omega )=\frac{8e^2Af_{\gamma N\Delta }^2M^2}
{27\pi^2m^6p_F^3}\,\frac{\omega ^2}{(\omega -\Delta)^2+\Gamma ^2/4}
\cdot\frac{p^4}{(p^2+m^2)^2}\,
        \label{app14}
\end{equation}
and
\begin{equation}
G^{NR}(p,\omega )=\frac{2e^2AM^2}{\pi ^2m^4p_F^3}
\,\left[1+\frac{m^4}{(p^2+m^2)^2}\right]
\frac{p^2}{(p^2+m^2)^2}
        \label{app15}
\end{equation}
respectively.
In eq.\,(\ref{app10}) the integration limits are
\begin{displaymath}
    L_1(Q)=Q-p_F\;,\quad L_2(Q)=\sqrt {2Q^2-p_F^2}-p_F
\end{displaymath}
and
\begin{displaymath}
        L_3(Q)=\sqrt {Q^2-p_F^2}  .
\end{displaymath}

\newpage
\section*{Figure Captions}
        \begin{abb}
        \protect\label{figA}
            Notation for three-momenta of the external particles a) for the
            one-nucleon process and b) for the two-nucleon reaction. The wavy
            lines denote photons and dashed lines denote pions. A circle indicates
            a bound, an arrow a free nucleon.
        \end{abb}

        \begin{abb}
        \protect\label{figB}
            Diagrammatical forms of the resonant part $T_i^{(R)}$ (first two
            terms) and the
            non-resonant part $T_i^{(NR)}$ (last term) of the amplitude $T_i$. These
            contributions to the $\gamma \pi NN$-interaction enter into the
            diagrams shown in Fig.\,(\ref{figA}).
        \end{abb}

        \begin{abb}
        \protect\label{figAA}
            Examples of diagrams, which have not been considered in this
            approach (cf. discussion in the text).
        \end{abb}

        \begin{abb}
        \protect\label{figC}
            Comparison of the calculated curve $\sigma (\omega )/A$ for nuclear
            photoabsorption with the experimental data. The dotted curve is the
            one-nucleon contribution $\sigma _1(\omega )$, while the dashed curve
            represents the two-nucleon mechanism $\sigma _1(\omega )$. The data
            are taken from Ref. \cite{ahrens}. The empty (full) circles correspond to 
            $^{208}$Pb ($^{12}$C) data, while the squares represent data 
            on $^{16}$O.
        \end{abb}

        \begin{abb}
        \protect\label{figD}
            Resonant (dashed) and non-resonant (dotted) contribution to the
            one-nucleon reaction (full line). The corresponding analytical
            expressions $\sigma _1^{(R)}(\omega )$, $\sigma _1^{(NR)}(\omega )$ and
            $\sigma _1(\omega )$, respectively, can be found in the Appendix.
        \end{abb}

        \begin{abb}
        \protect\label{figE}
            Same as Fig.\,(\ref{figD}), but for the two-nucleon reaction.
            The analytical
            expressions $\sigma _2^{(R)}(\omega )$, $\sigma _2^{(NR)}(\omega )$ and
            $\sigma _2(\omega )$, are also given in the Appendix.
        \end{abb}

        \begin{abb}
        \protect\label{figH}
            Contribution of interference terms. The full (dashed) curve
            corresponds to the one- (two-) nucleon case. As can be seen from the
            overall scale, both for $\sigma _1$ and $\sigma _2$ this
            contribution is highly suppressed.
        \end{abb}

        \begin{abb}
        \protect\label{figF}
            Effect of corrections of the order $1/M$ in the one-nucleon
            mechanism. The full (dashed) curve is with (without) relativistic
            corrections.
        \end{abb}

        \begin{abb}
        \protect\label{figJ}
            Correlation function $S_C$. The dashed curve corresponds to $S_0$,
            while in the dotted curve further two-nucleon correlations and
            three-nucleon correlations have been included.
        \end{abb}

        \begin{abb}
        \protect\label{figG}
            Effect of two- and three-nucleon correlation in the case of the
            non-relativistic one-nucleon part of the photoabsorption cross section.
            The dashed curve contains only $S_0$, while in the full curve
            $S_C$ (cf. Fig.\,(\ref{figJ})) has been included.
        \end{abb}

        \begin{abb}
        \protect\label{figI}
            Approximate description of differential cross sections for helium.
            The differential cross section with respect to the direction of the
            outgoing proton is shown as a function of the energy of the incoming
            photon. Data points are taken from \cite{wich}. The full curve and
            filled circles correspond to the one-nucleon process, while the dashed
            curve and empty circles represent the two-nucleon case.
        \end{abb}

\newpage

\begin{figure}[t]
\begin{center}
    \caption{}
    \epsfig{figure=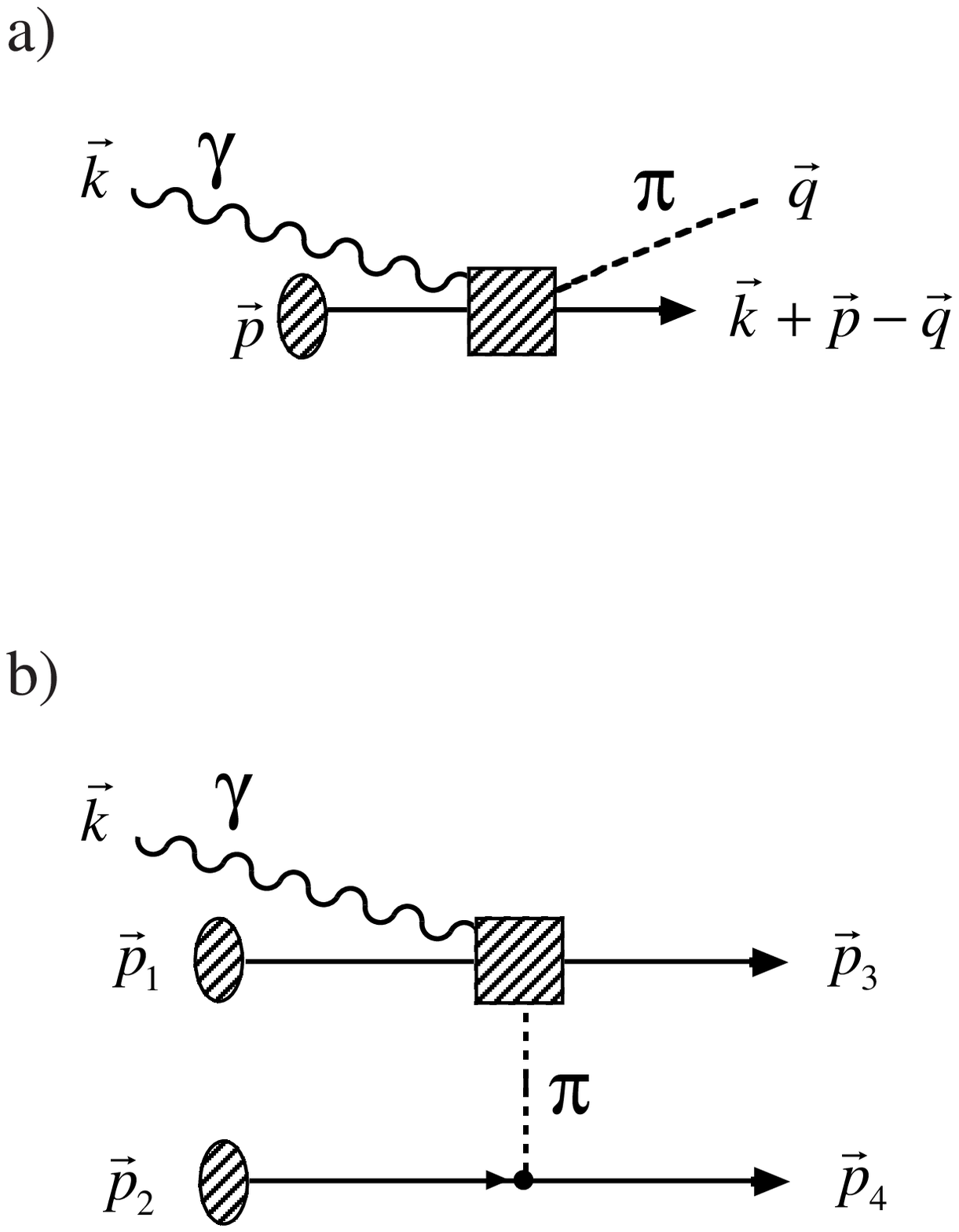,width=14cm}\\
\end{center}
\end{figure}

\begin{figure}[t]
\begin{center}
    \caption{}
    \vspace{1cm}
    \epsfig{figure=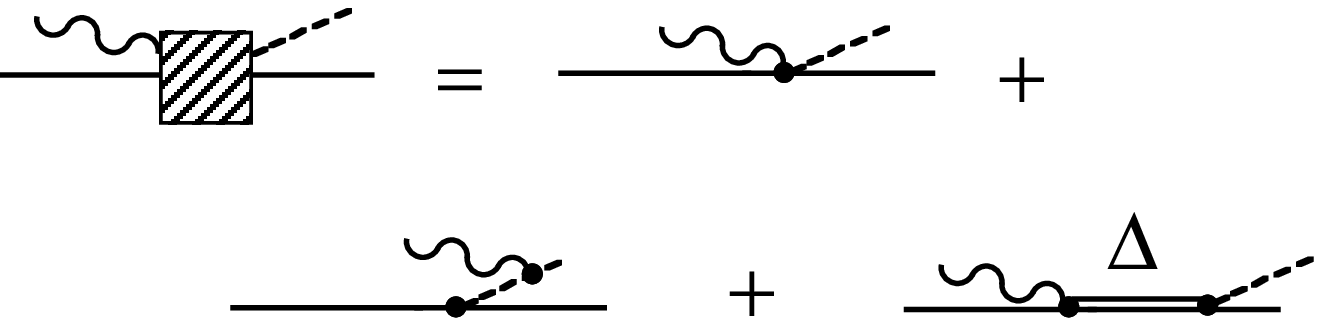,width=16cm}\\
\end{center}
\end{figure}

\begin{figure}[t]
\begin{center}
    \caption{}
    \vspace{1cm}
    \epsfig{figure=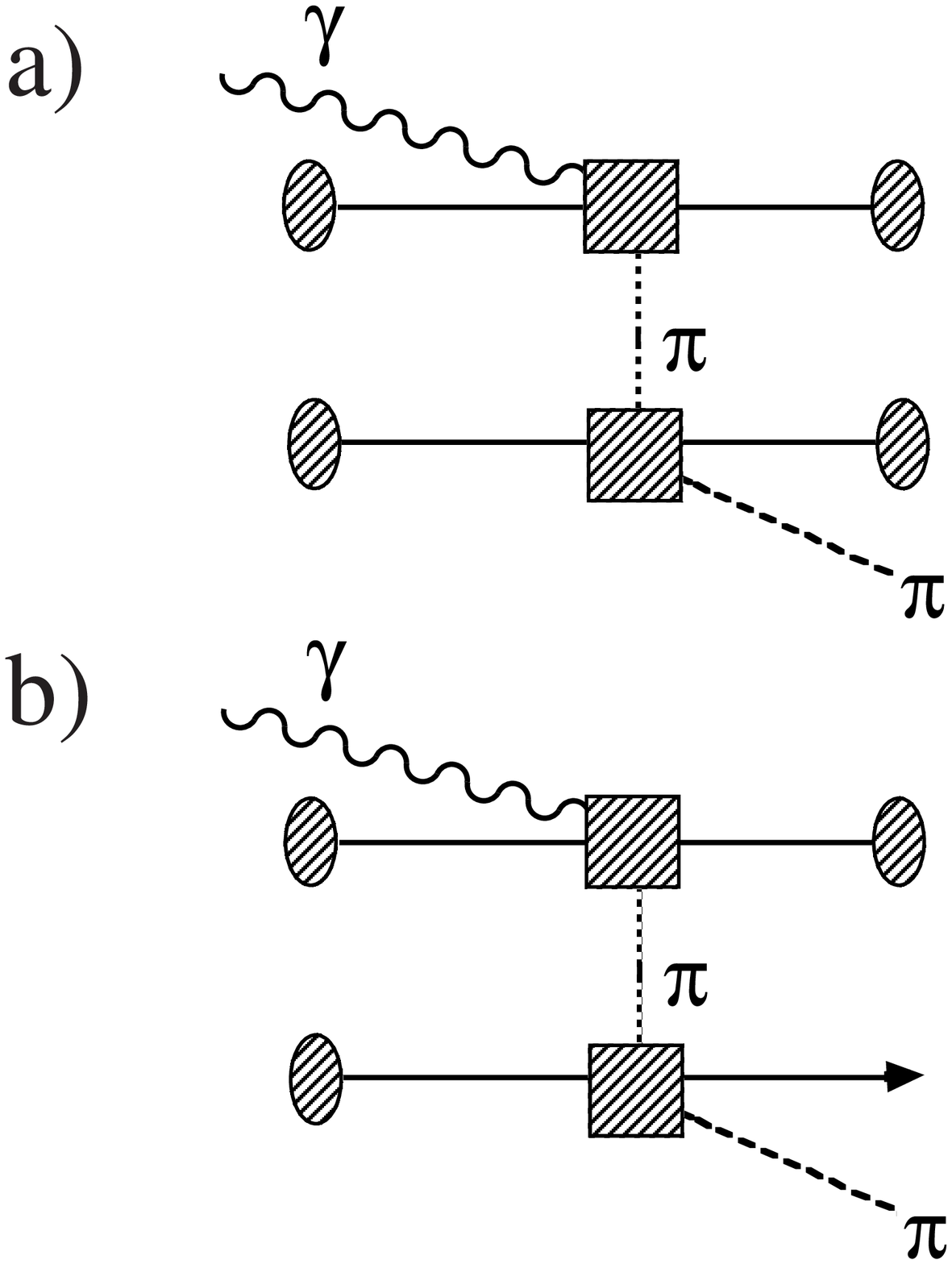,width=10cm}\\
\end{center}
\end{figure}

\begin{figure}[t]
\begin{center}
    \caption{}
    \vspace{1cm}
    \epsfig{figure=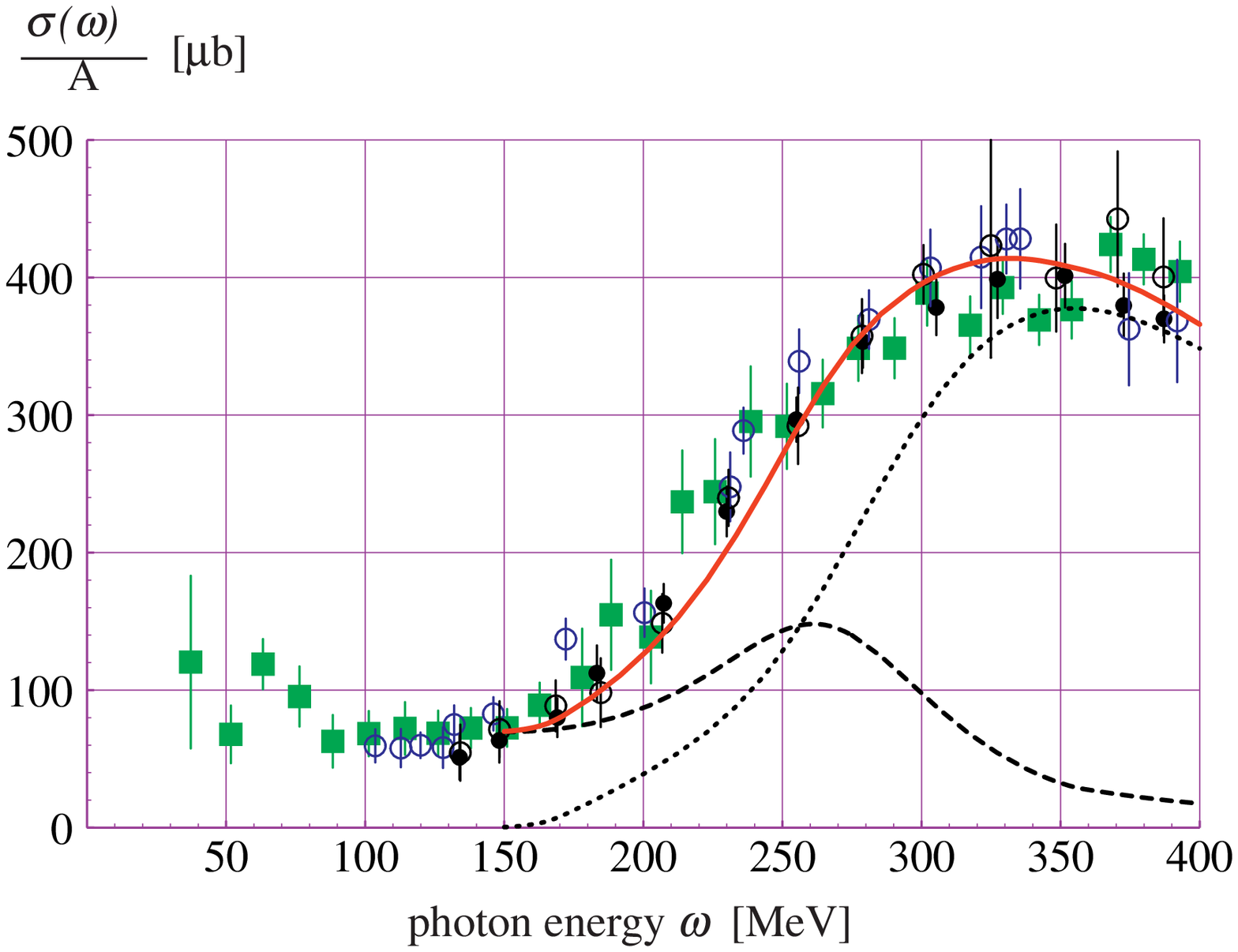,width=16cm}\\
\end{center}
\end{figure}

\begin{figure}[t]
\begin{center}
    \caption{}
    \vspace{1cm}
    \epsfig{figure=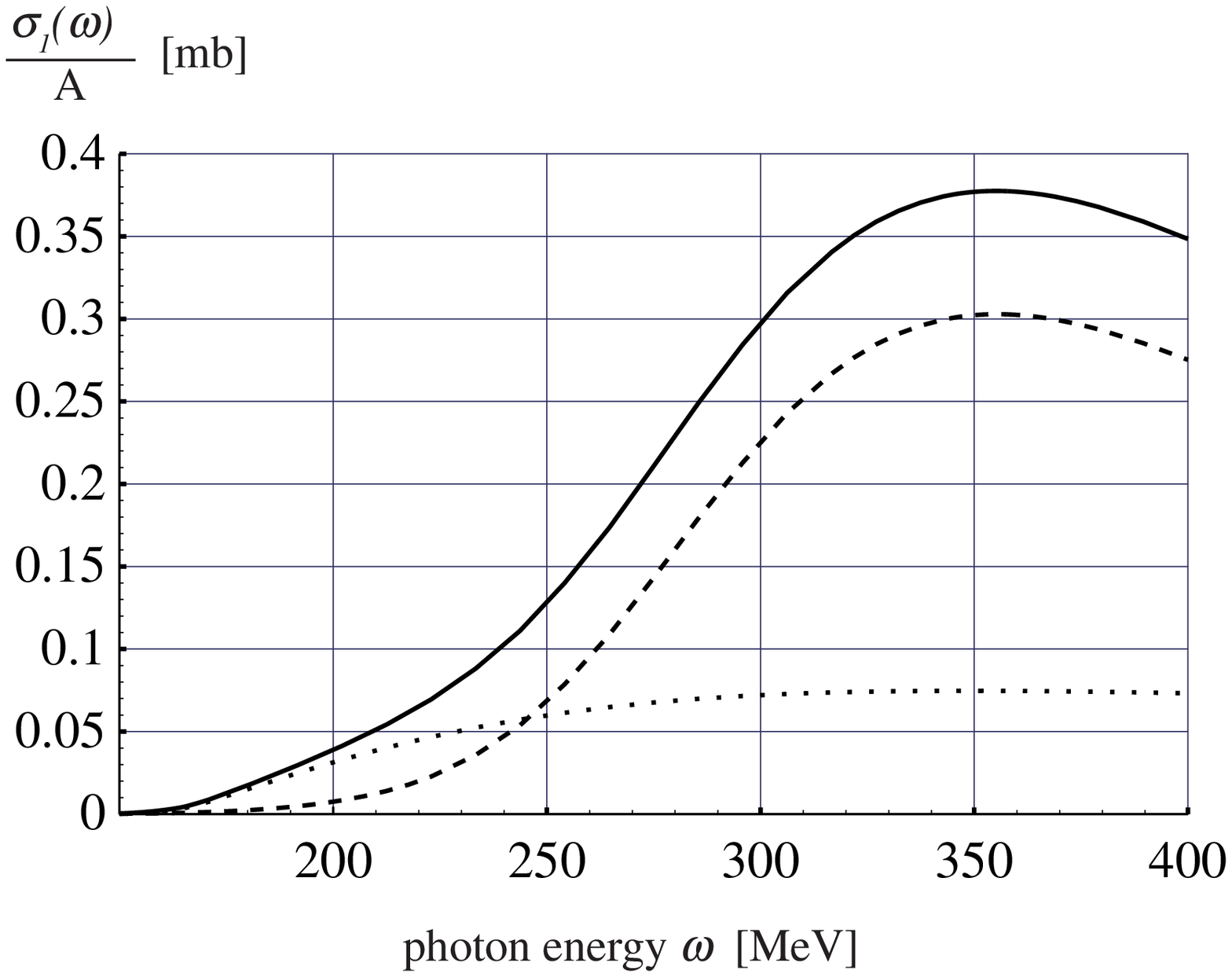,width=16cm}
\end{center}
\end{figure}

\begin{figure}[t]
\begin{center}
    \caption{}
    \vspace{1cm}
    \epsfig{figure=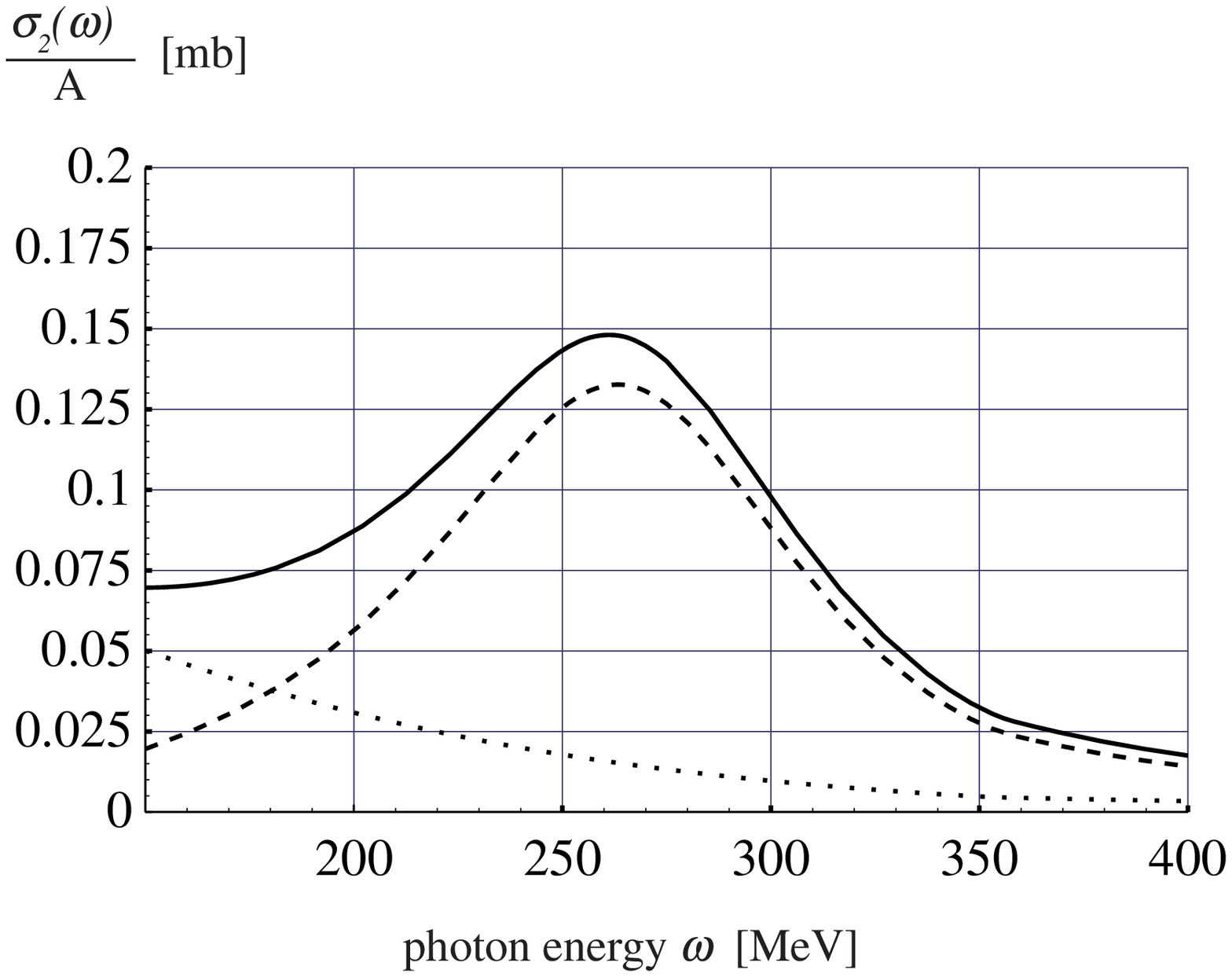,width=16cm}
\end{center}
\end{figure}

\begin{figure}[t]
\begin{center}
    \caption{}
    \vspace{1cm}
    \epsfig{figure=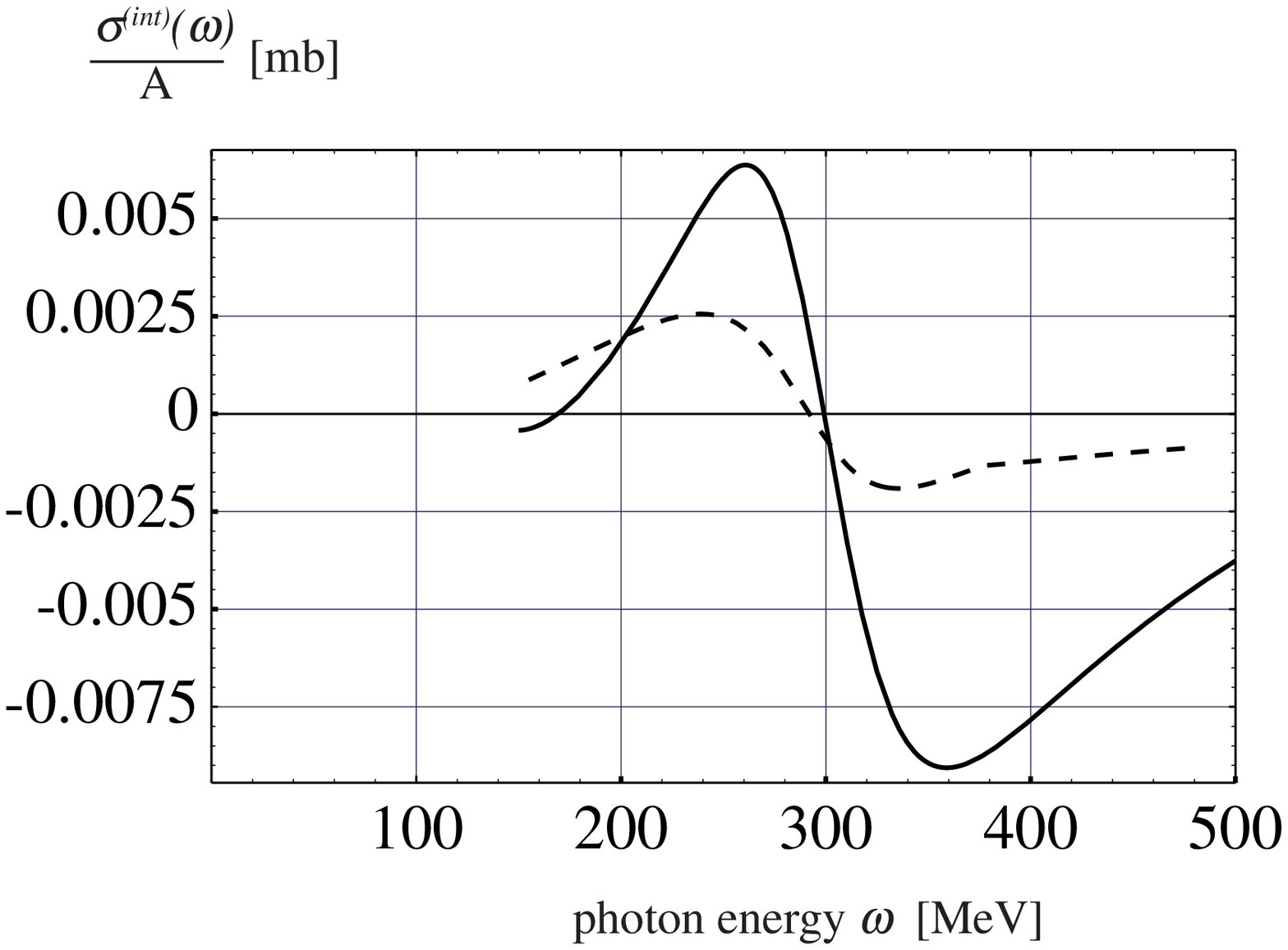,width=16cm}
\end{center}
\end{figure}

\begin{figure}[t]
\begin{center}
    \caption{}
    \vspace{1cm}
    \epsfig{figure=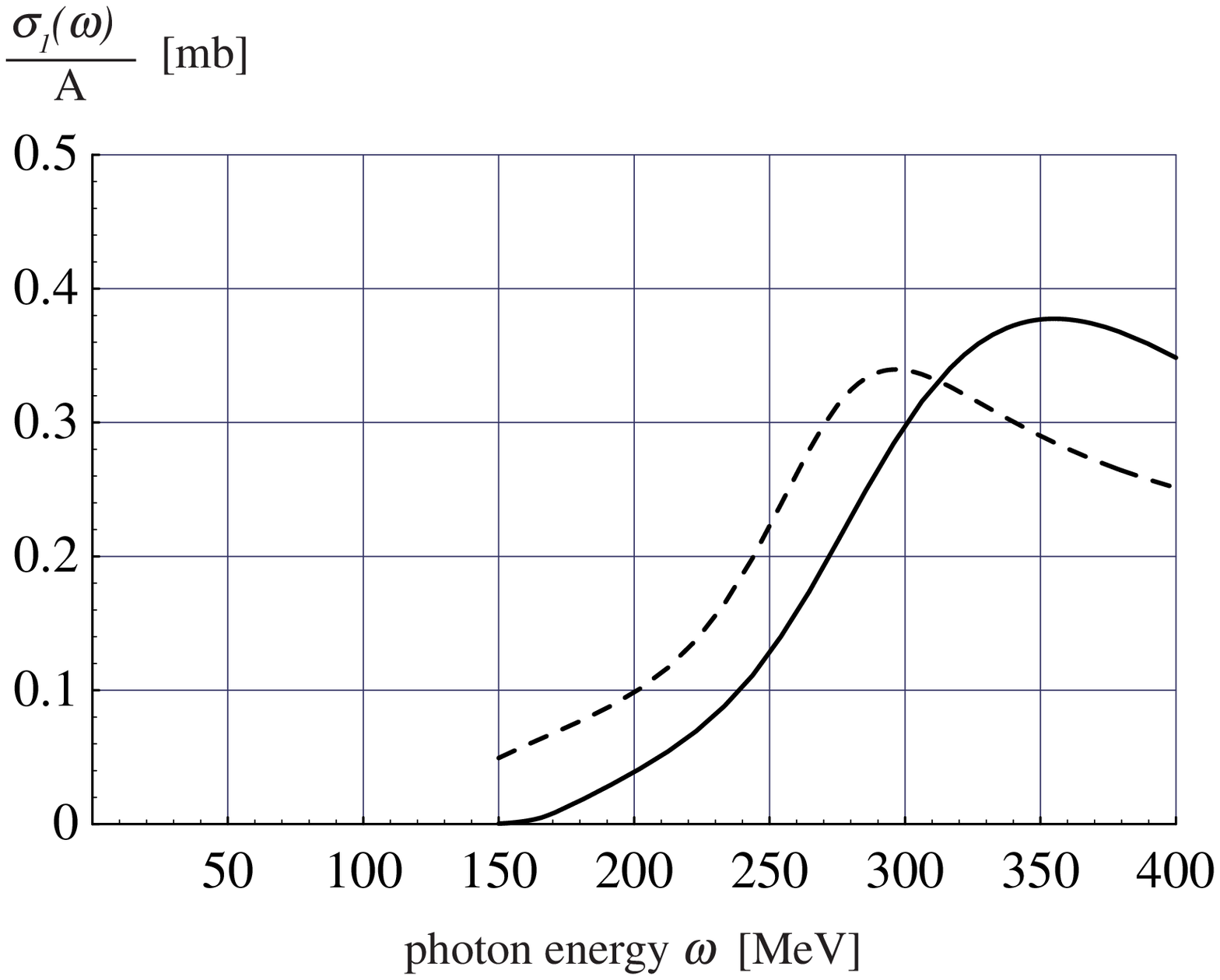,width=16cm}
\end{center}
\end{figure}

\begin{figure}[t]
\begin{center}
    \caption{}
    \vspace{1cm}
    \epsfig{figure=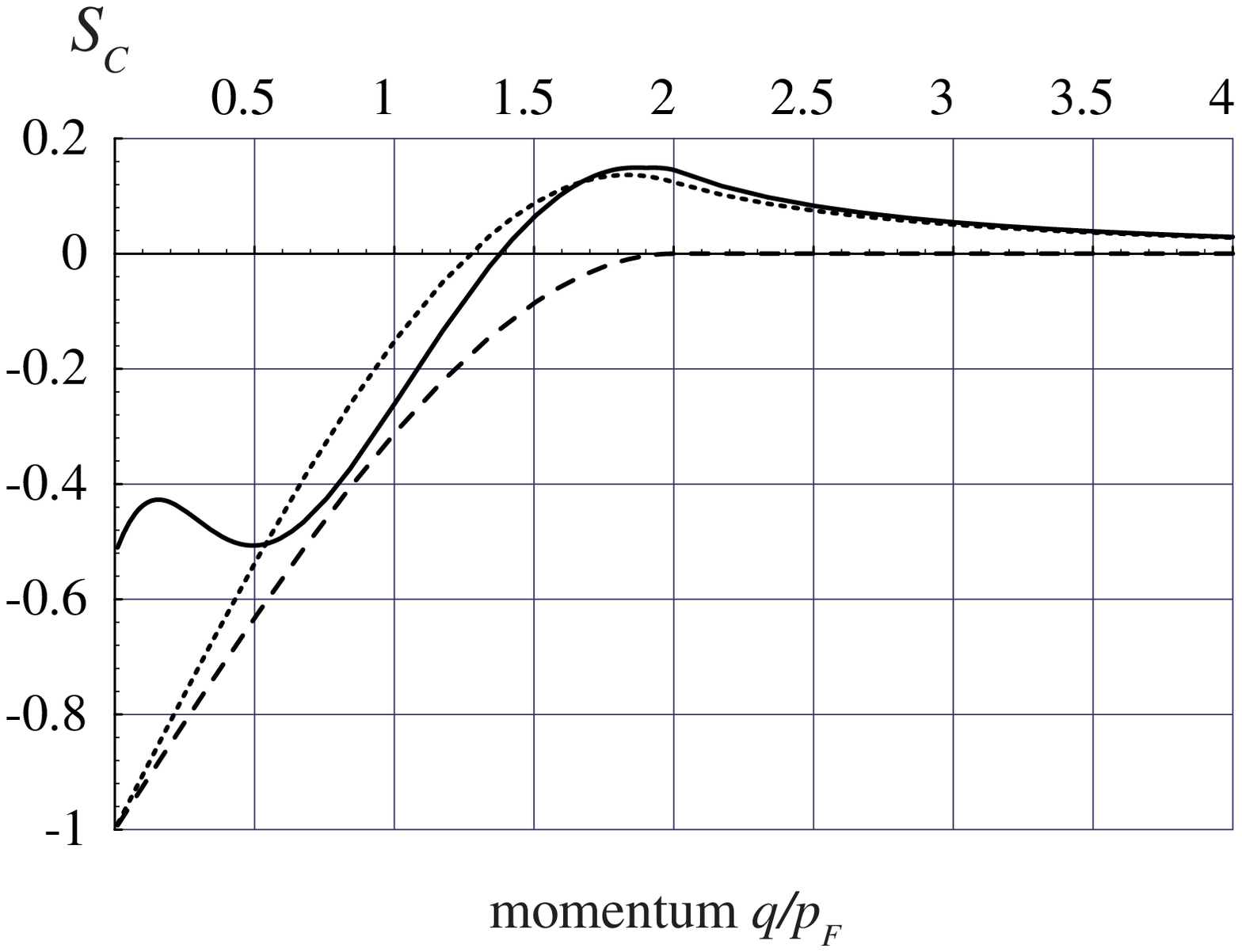,width=14cm}
\end{center}
\end{figure}

\begin{figure}[t]
\begin{center}
    \caption{}
    \vspace{1cm}
    \epsfig{figure=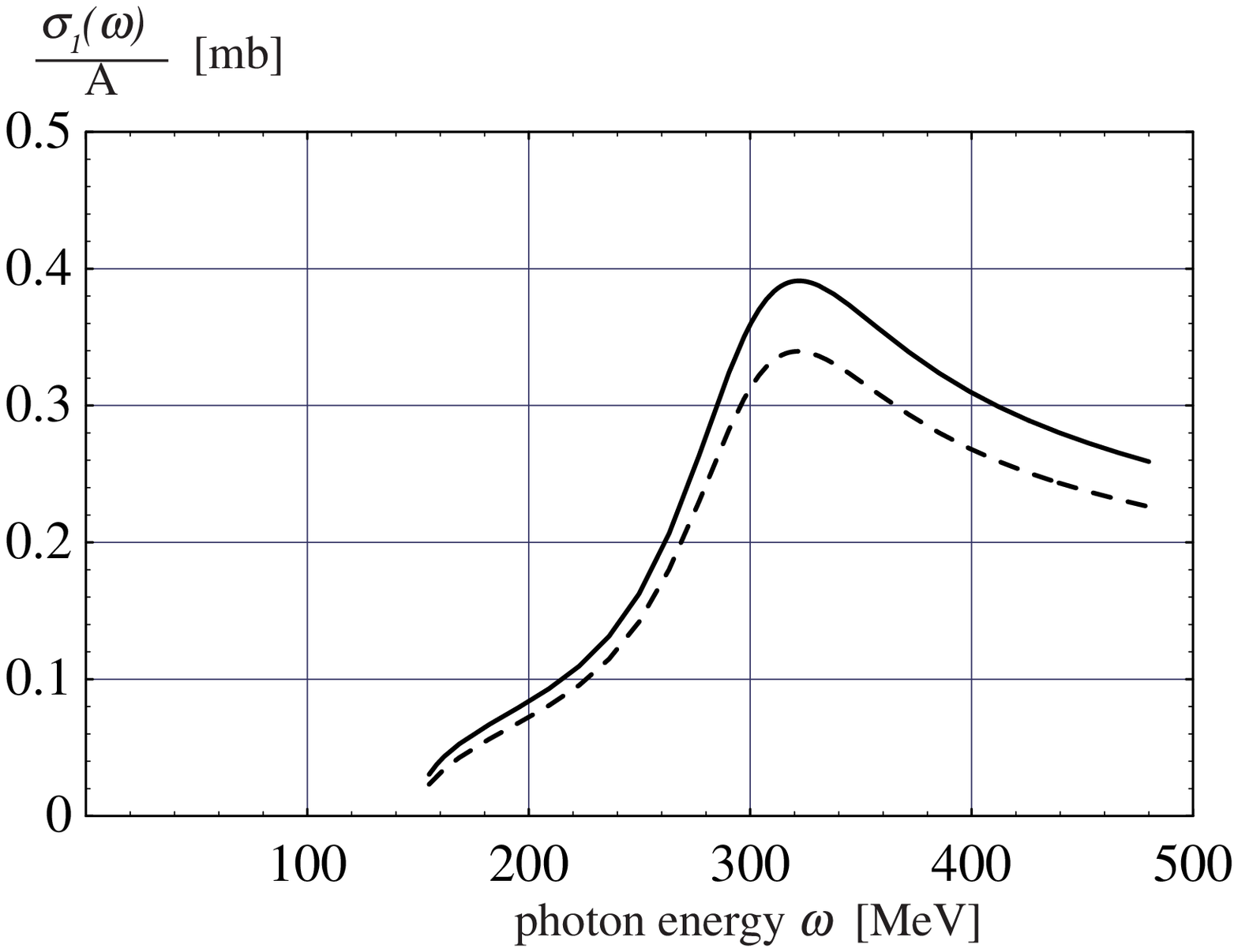,width=16cm}
\end{center}
\end{figure}

\begin{figure}[t]
\begin{center}
    \caption{}
    \vspace{1cm}
    \epsfig{figure=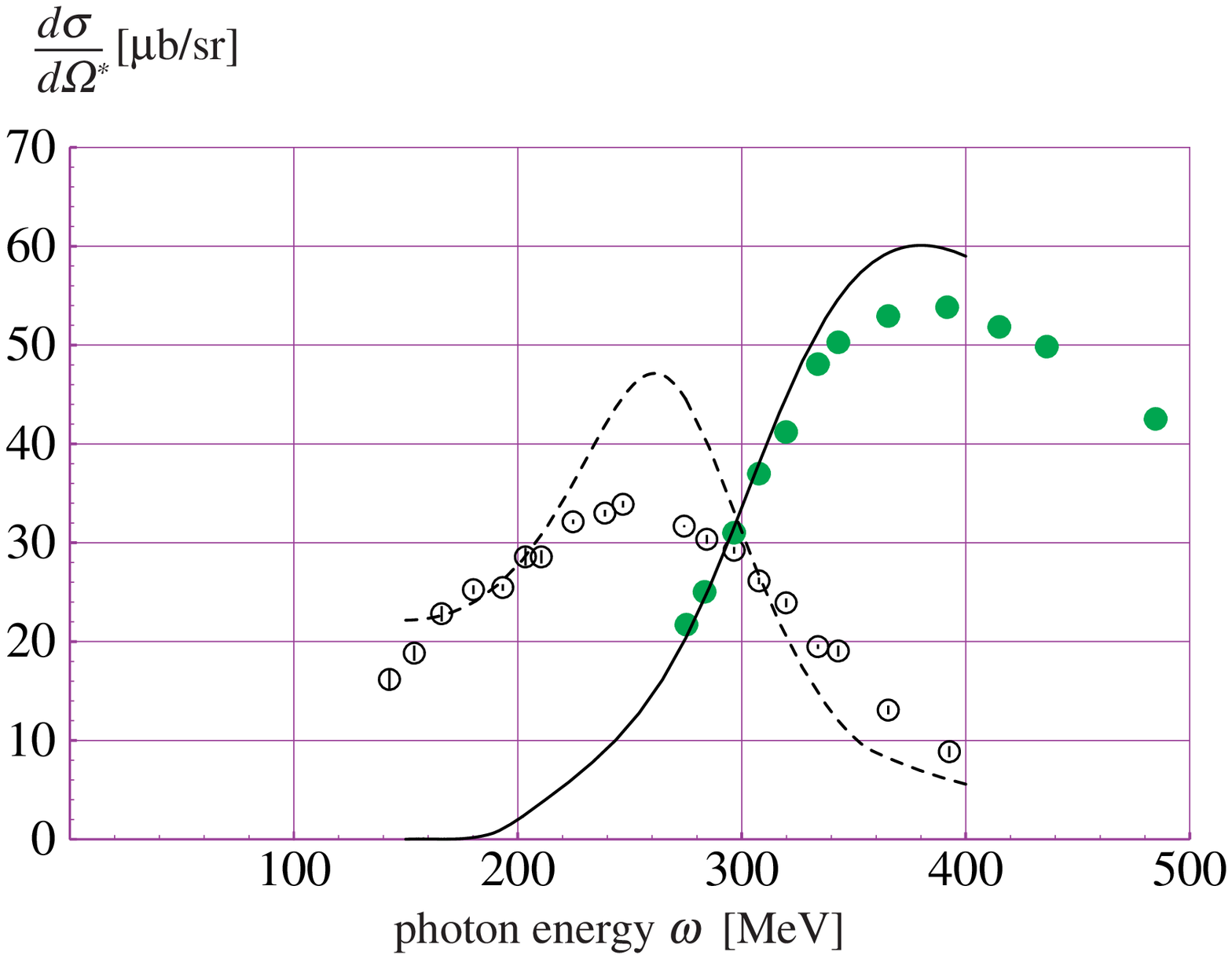,width=14cm}
\end{center}
\end{figure}

\end{document}